\documentstyle[epsfig]{aipproc}
\begin{document}

\title{Galactic Diffuse $\gamma$-ray Emission at TeV Energies
and the Ultra-High Energy Cosmic Rays}

\author{G. A. Medina Tanco$^{*, \dagger}$ and 
E. M. de Gouveia Dal Pino$^{*}$ }

\address{$^*$
Instituto Astron\^{o}mico e Geof\'{\i}sico, Universidade de S\~{a}o Paulo 
\thanks{This work was partially sponsored by the Brazilian Agencies FAPESP 
and CNPq}, (04301-904) S\~{a}o Paulo - SP, Brasil \\
$^{\dagger}$ Royal Greenwich Observatory, 
Cambridge CB3 0EZ,
UK, \\ 
gmt@ast.cam.ac.uk}


\maketitle

\begin{abstract}
Using the cosmic ray  (CR) data available in the energy interval $(10 -
2 \times 10^{7})$ GeV/particle,
we have calculated the
profile of the primary $\gamma$-ray spectrum produced 
by the interaction of  these CR with thermal nuclei 
of the ISM. Normalized to the EGRET measurements, this 
allows an estimate of the galactic diffuse $\gamma$-ray background
due to intermediate and high energy CR at TeV energies.
On the other hand, over the last few years,
several particles with energies above 
$10^{20}$ eV (beyond the Greisen-Zatsepin-Kuzmin cut-off)
have been detected. These particles are very likely extragalactic
protons originated at distances not greater than 
$30 - 50$ Mpc [e.g., 1].
The propagation of these ultra-high energy protons (UHEP) through
the intergalactic medium leads to the development of
$\gamma$-ray cascades and an ultimate signature at TeV energies.
To assess the statistical significance of this $\gamma$-ray
signature by the UHEP, we have also simulated the development 
of electromagnetic cascades triggered by the decay of a
$10^{19}$ eV $\pi^{o}$ in the intergalactic medium after 
an UHEP collision with a cosmic microwave background photon.  
 
\end{abstract}

\section*{The $\gamma$-Ray Spectrum at $10^{12} - 10^{15}$ eV}

%
%
%
%

The $\gamma$-ray production mechanisms related to CR interactions 
with the ISM are well understood and have been described in
detail by a number of authors [e.g., 2]. The processes 
which contribute to diffuse $\gamma$-ray production are: 
(i) bremsstrahlung; (ii) inverse Compton scattering; 
and (iii) nuclear interactions,
but for the  energies of 
interest in the present work, which are above 10 GeV,
the latter is the most relevant mechanism. 
 
In order to calculate the $\gamma$-ray differential spectrum, 
$dN_{\gamma}/dE$, 
from the observed CR differential spectrum, 
$dJ_{cr}/dE$, we have employed the {\it Lund 
Monte Carlo for Hadronic Processes}  routines  
(PYTHIA version 6.1, March 1997) [3].
The CR energy spectrum includes data from 
BASJE, TIBET, TACT, TUNKA, JACEE and Grigorov [4].
CR nuclei interact with the ISM thermal nuclei producing 
$\gamma$-ray through different processes (e.g., 
$\pi^{o} \rightarrow 2\gamma $;
$\pi^{+} \rightarrow \mu^{+} \rightarrow e^{+} \rightarrow \gamma$;
$q\bar q \rightarrow g\gamma$;
$f\bar f \rightarrow \gamma \gamma$;
$qg \rightarrow q\gamma$;
etc). 
The resulting $\gamma$-ray spectrum is depicted in Fig. 1.

\begin{figure}[t!] 
\vspace{-1.0cm}
\centerline{\epsfig{file=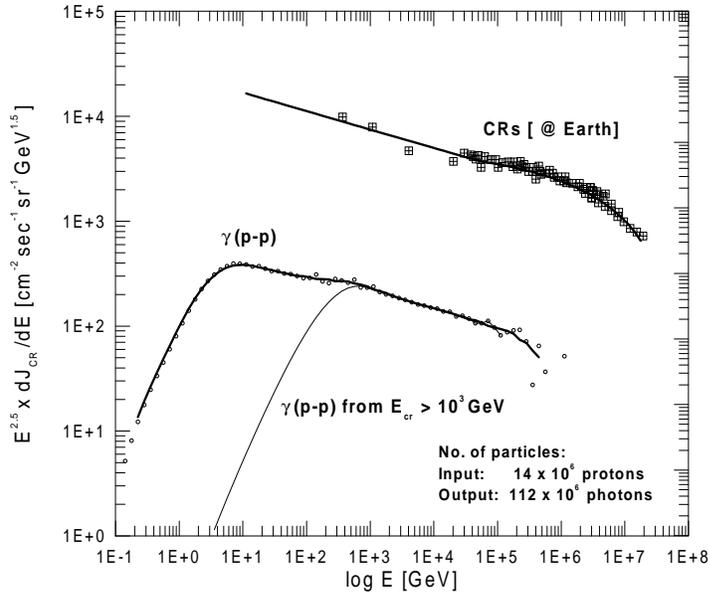,height=15.0cm,width=10.0cm}}
\vspace{-4.0cm}
\caption{
Observed CR spectrum 
(obtained from BASJE, TIBET, TACT, TUNKA, 
JACEE and Grigorov experiments), and the calculated 
background  $\gamma$-ray spectrum at TeV energies. }
\vspace*{10pt}
\label{fig1}
\end{figure}

\begin{figure}[t!] 
\vspace{-1.0cm}
\centerline{\epsfig{file=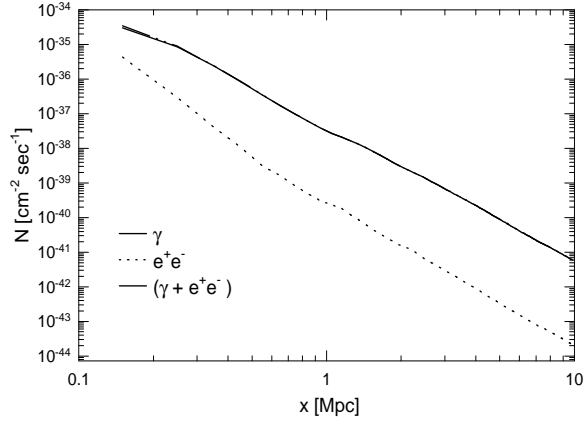,height=15.0cm,width=10.0cm}}
\vspace{-5.5cm}
\caption{
Total flux of $\gamma$-ray photons and
 electron-positron pairs [$cm^{-2}$ $s^{-1}$]
produced in a cascade triggered by a $10^{20}$ eV 
proton as a function of the distance along 
the axes of the cascade ($x$).
}
\vspace*{10pt}
\label{fig2}
\end{figure}

\begin{figure}[t!] 
\vspace{-1.2cm}
\centerline{\epsfig{file=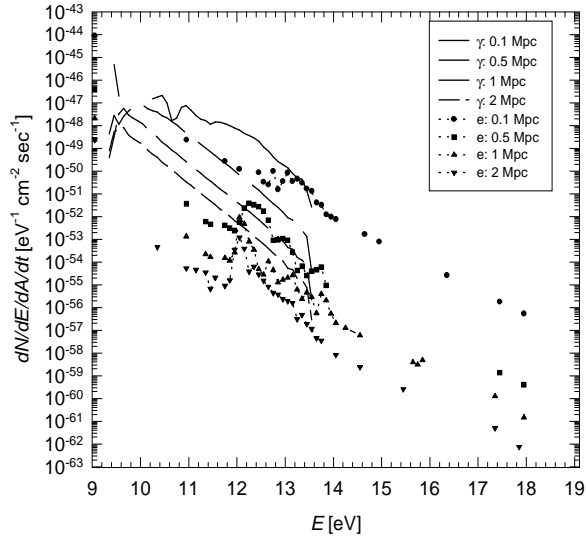,height=15.0cm,width=10.0cm}}
\vspace{-4.5cm}
\caption{
Spectra of $\gamma$-ray photons and electrons 
[$eV^{-1}$ $s^{-1}$ $cm^{-2}$]
produced in a cascade triggered by a $10^{20}$ eV-proton 
for different  distances $x$.
 }
\vspace*{10pt}
\label{fig3}
\end{figure}

\begin{figure}[t!] 
\vspace{-1.5cm}
\centerline{\epsfig{file=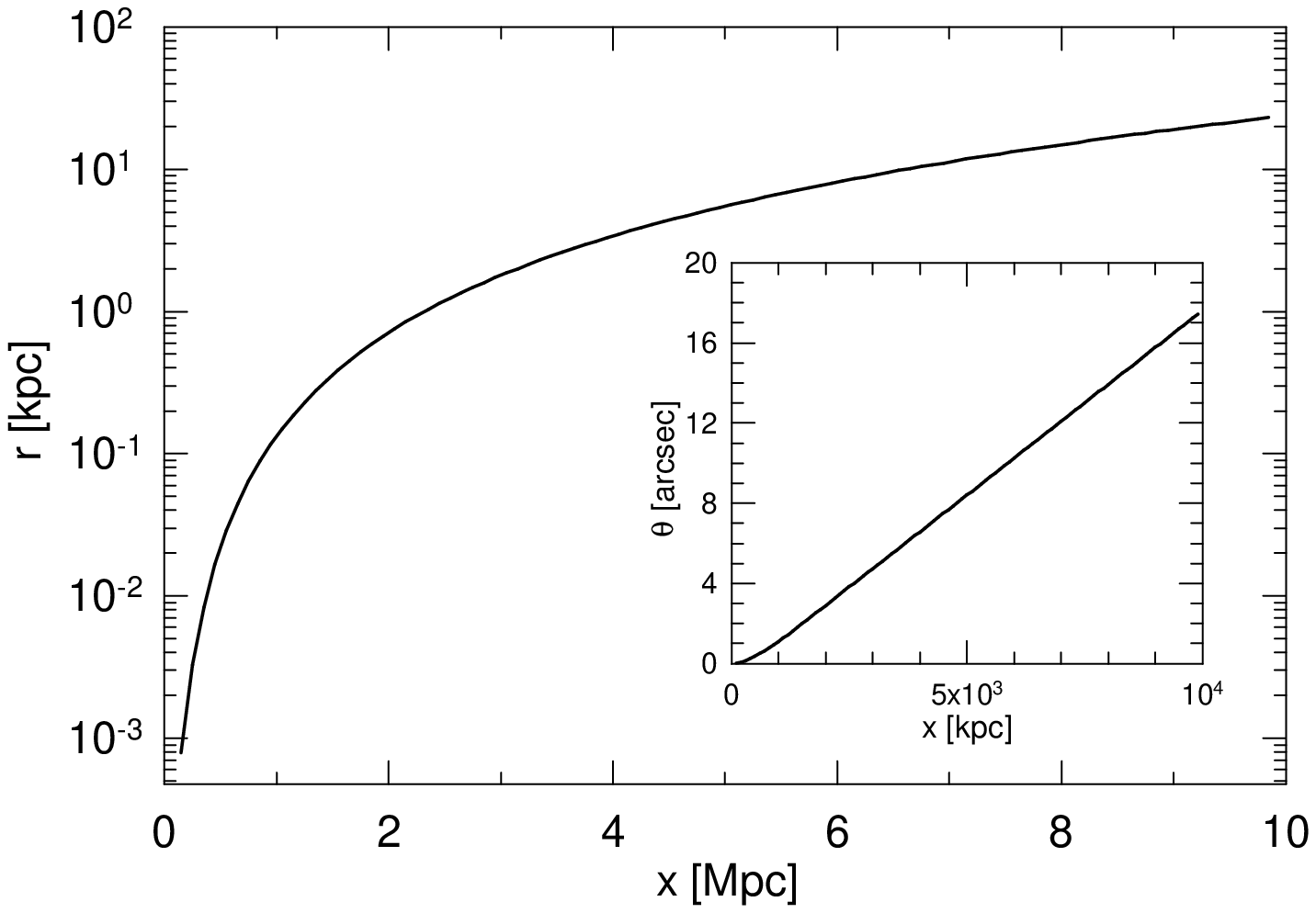,height=15.0cm,width=10.0cm}}
\vspace{-5.0cm}
\caption{
Radius ($r$) and aperture angle ($\theta$) 
of the cascade in the laboratory reference frame as a function
of the distance along the cascade axes ($x$).
 }
\vspace*{10pt}
\label{fig4}
\end{figure}


The evaluation of the  mean-free path of
the $\gamma$-ray photons through the galactic background radiation field
(stellar, IR, and cosmic microwave background (CMB) photons) shows that
the primary $\gamma$-spectrum at
$E_{\gamma} \simeq (10^{10} - 10^{15} )$ eV 
is almost unaffected by interactions 
with background photons [5].  
Therefore, the actual value of the galactic $\gamma$-background 
at these energies can be estimated by normalizing the $\gamma$-spectrum
of figure 1 to the EGRET observations at 1-10 GeV.

\section*{Electromagnetic Cascades due to UHEP}
 
For the analysis of the electromagnetic cascading triggered
by UHEP propagation through 
the intergalactic medium (IGM), two different situations must 
be considered: (i) when an intergalactic magnetic field
is present ($B_{IGM} \not= 0$);  or (ii) when it is absent
(e.g., inside the voids) 
($B_{IGM} \simeq 0$). In the first case, synchrotron
radiation will
prevent the development of cascading due to 
the rapid draining of energy of the secondary 
electrons into low energy photons

\begin{equation}
E_{\gamma} \approx 2 \times 10^{11}  \left(\frac{B_{\perp} } {10^{-9} G}\right) \left(\frac {E_e} {10^{20} eV}\right)^2 eV
\end{equation}

Moreover, for the $\approx 10^{8}$ photons produced,
 relativistic beaming and curvature of the 
electron's trajectory will reduce to only 
$\sim  2 \times 10^{-3}$ the number of photons
that can be detected per event, per observer
[5].
Thus, we must focus on the case for which  
$B_{IGM} \simeq 0$.
In this case, the cascade is initiated by the 
interaction of an UHEP with a photon of the CMBR. 
Either a neutral or a charged pion may be produced
and the initial energy is
subsequently channeled to lower and lower energies through a 
$\gamma\gamma$-pair production, inverse Compton cycle [6],
until no further $\gamma$-ray is 
produced when then the threshold for pair production due to
interactions with the CMBR photons is reached. 
Thereafter, the interactions must involve higher background 
photons and the corresponding mean-free path,
 $\lambda_{\gamma \gamma}$, increases rapidly 
[see, e.g., 7-8]. 

As an example, Figs. 2-4 show the results for a typical cascade triggered 
by a $10^{20}$ eV UHEP while traversing a very low magnetic field 
($B \lesssim 10^{-12}$ G) region of the IGM.

\section*{Conclusions and Discussion}

The previous results indicate that, as soon as the shower effectively 
develops (i.e., at $d \gtrsim 100$ kpc, from the $\pi^{o}$ decay), 
low energy electrons are produced.
Consequently, an upper limit for the duration of the cascade at the
detector can be calculated as the time delay 
 between a $10^{12}$ eV electron and a $\gamma$-ray photon
[$\Delta t \approx (1 - \beta) d/c$]:

\begin{equation}
\Delta t \simeq 0.1 \left( \frac{d}{1 Mpc}\right) s 
\end{equation}

The results above also indicate that a 
cascade initiated by a single UHEP interaction 
with CMB photons should reduce to, at most, 
one photon of relatively low energy at the 
detector (see [5] for details). 
This renders the UHEP-CMB photon interaction practically unobservable.

The situation turns out to be different 
if one considers the background of
$\gamma$-rays produced by the whole distribution 
of UHEP interacting with the CMBR

\begin{equation}
F_{obs}(r) = \frac{\pi \nu_{p,\gamma}}{4} \int^r_0   \theta(r)^2 \Phi(r) 
\Delta t(r) r^{2} dr   
\end{equation}

\noindent
Where $\nu_{p,\gamma}$ is the number of  UHEP-$\gamma_{CMB}$ 
interactions per unit time, per unit volume. 
Since the estimated flux of UHECR is 
$J(E>10^{20})$ eV) 
$\approx 3.3 \times 10^{-21}$ $cm^{-2}$ $sr^{-1}$ $s^{-1}$
 [9], $\nu_{p,\gamma} \approx 10^{-45} cm^{-3} s^{-1}$. 
Considering the volume of the local Universe within which the
 UHECR sources must be located ($d \lesssim 50 $ Mpc),
 we derive a lower limit for the contribution of UHECR to 
the background diffuse $\gamma$-ray  flux 
$F_{obs} \approx 10^{-10} $ $cm^{-2}$ $s^{-1}$. 
This value is about an order of magnitude smaller than the galactic
 diffuse background as estimated from our calculations and
 from the EGRET data [5]. 
Nonetheless, this contribution is comparable to the $\gamma$-ray
 diffuse background component due to blazers (see, e.g., 
 Fegan, this conference [10]).

\end{document}